\documentstyle[psfig,conf_iap,]{article}
\begin{document}
\heading{%
%Begin Heading
%
The Distribution of Thermal Pressures\\
in the Interstellar Medium
%
%End Heading
} 
\par\medskip\noindent
\author{%
%Begin Author names
Edward B. Jenkins$^{1}$, Todd M. Tripp$^{1}$
%End Author names
}
\address{%
%First address
Princeton University Observatory, Princeton, NJ 08544-1001, USA
}

\begin{abstract}
It is generally recognized that the interstellar medium has a
vast range of densities and temperatures.  While these two
properties are usually anticorrelated with each other, there
are nevertheless variations in their product, $nT$, the thermal
gas pressure (divided by the Boltzmann constant $k$).  A study
of these variations and their kinematical relationships can help to mold
our concepts of interstellar gas dynamics and give us insights
on secondary effects that might arise from turbulence.

In neutral gas, the relative populations of neutral carbon
atoms in the excited fine-structure states can give a direct
measure of a local thermal pressure.  A picture of the
distribution function for thermal pressures in H I regions is
now arising from a survey of interstellar C~I absorption features
in the UV spectra of 21 early-type stars, observed with a wavelength resolving power of
200,000 by the STIS instrument on the Hubble Space Telescope.
Most of the gas is within the range $10^3 < p/k < 10^4\,{\rm cm}^{-3}$
K, but there is also evidence for some of the material being
at much higher pressures, i.e., $p/k > 10^5\,{\rm cm}^{-3}$K.  While
the fraction of gas at these elevated pressures is quite small,
it seems nearly ubiquitous.  This phenomenon may arise from
small-scale, short-lived density enhancements that are produced
by converging flows of material in supersonic turbulence.

\end{abstract}
\section{Background}
Constituents of the interstellar medium are viewed in projection, and
thus we measure the integral of a density (or square of the density)
along a finite path, or one of indefinite length, depending on the
character of the observation.  To obtain explicit indications of local
particle densities in individual regions, we must examine ratios of
certain constituents, such as the ionized and neutral forms of atoms,
H$_2$ rotational populations, and atomic excited fine-structure
excitations.  We concentrate here on the last of these three
diagnostics.  To study fine-structure level populations, one must
observe absorption lines in the ultraviolet.  The best prospects are
from transitions arising from the singly-charged ions of N, Si and C or
the neutral forms of O and C.  (For a comprehensive bibliography on
collision rate constants, see [1]).  C~I is a good probe of
conditions in H~I regions, and it is relatively easy to excite by
collisions.  However since carbon is mostly singly ionized, its neutral
form emphasizes regions where either the photo-ionization rate $\Gamma$
is reduced or electron densities $n(e)$ are high.

\begin{figure}\label{f1f2_diag}
\centerline{\vbox{
\psfig{figure=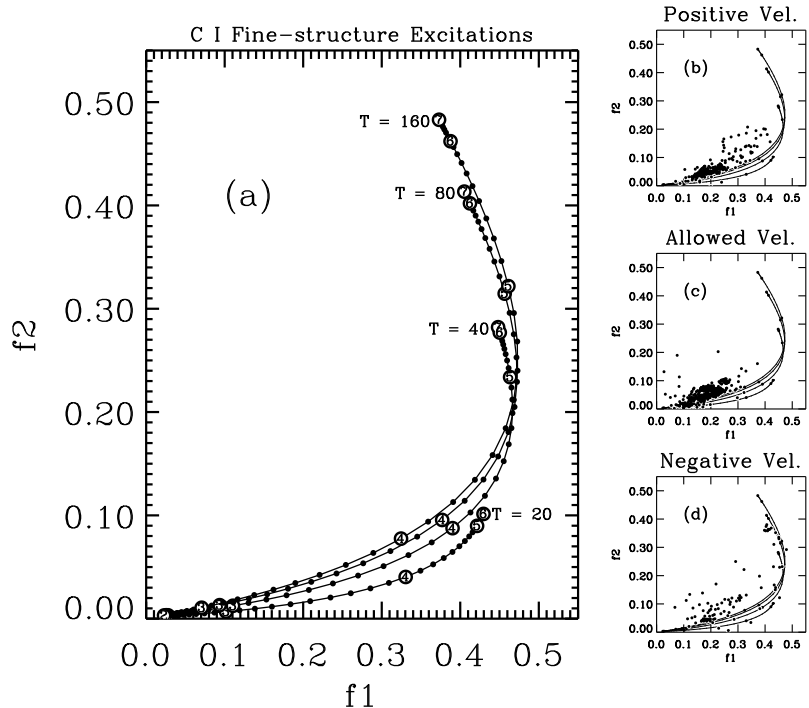,height=9.cm}
}}
\caption[]{$(a)$: Constant-temperature tracks of C~I excitation for
varying thermal pressures, with demarkations in the form of (circled)
integral values of $\log p/k$ (small dots represent increments of 0.1
dex). $(b,c,d)$: Observed outcomes for all 21 stars in the STIS survey,
with the C~I segregated into three kinematic categories (see text).}
\end{figure}

The ground electronic state of C~I is split into three levels, a
zero-energy fine-structure level, $^3{\rm P}_0$, and two excited levels,
$^3{\rm P}_1$ and $^3{\rm P}_2$ with thermal excitation energies
$E/k=23.6\,$K and 62.4$\,$K.  Following a usual convention, we refer to
C~I in these states as C~I, C~I$^*$, and C~I$^{**}$, respectively.  A
good way to express the observed population ratios is through two terms,
$f1\equiv N({\rm C~I}^*)/N({\rm C~I_{total}})$ and $f2\equiv N({\rm
C~I}^{**})/N({\rm C~I_{total}})$, where $N({\rm C~I_{total}})=N({\rm
C~I})+N({\rm C~I}^*)+N({\rm C~I}^{**})$.  These quantities have the
useful property that any superposition of contributions from different
regions, ones that are not resolved by shifts in Doppler velocity,
create a net combination of $f1$ and $f2$ that is at the C~I-weighted
``center of mass'' of points that could be attributed to individual
regions in a diagram of $f1$ {\it vs.\/} $f2$.  This is an important
property of this representation that we will use later in this paper. 
The tracks of $f1$ and $f2$ for homogeneous (pure atomic hydrogen)
regions at selected temperatures and varying thermal pressures, $nkT$,
is shown in panel $(a)$ of Fig.~\ref{f1f2_diag}.

\section{High-resolution survey of C~I using STIS}

To study in detail the absorption features of C~I, we observed 21
early-type stars with the {\it Space Telescope
Imaging Spectrograph\/} (STIS) on HST.  The E140H grating and a very narrow
entrance slit (0.03 arc-sec wide) allowed us to obtain the highest
resolution possible ($\lambda/\Delta\lambda = 200,000$ or $1.5\,{\rm
km~s}^{-1}$) [2].  To maximize our observing
efficiency, we selected mostly stars within the Continuous Viewing Zone
(CVZ) of HST, and this meant that most of our stars in the Galactic
plane had longitudes clustered near $\ell=120^o$ or 300$^o$.

Considering the fact that nearly all lines of sight have multiple
velocity components spread over tens of ${\rm km~s}^{-1}$, absorption
features from each transition within a multiplet usually overlap each
other.  To unravel these confusing blends, we took advantage of the fact
that different multiplets have different arrangements of lines, so by
observing many of them we could solve for the unique column density {\it
vs.\/} velocity profiles for  C~I, C~I$^*$, and C~I$^{**}$ that gave the
best agreement with the observations.  We accomplished this by defining
a system of 300 simultaneous linear equations (for the apparent optical
depths of three species for 100 velocity channels) that would solve for
the minimum $\chi^2$ over 9 different multiplets.

The results of the survey are shown in the small panels of
Fig.~\ref{f1f2_diag}.  Each dot in one of the panels represents
$N=10^{13}\,{\rm cm}^{-2}$ of C~I.  The outcomes are segregated
according to where the radial velocity is in relation to an expected
``allowed'' velocity range spanned by $v_{lsr}=0$ and the computed
velocity for differential Galactic rotation at the position of the
target star.  Results obtained above $(b)$ and below $(d)$ this range
represent gas with peculiar motions, while gas within the range $(c)$
represents mostly quiescent material (and some disturbed material moving
transverse to the line of sight).  Differences in the distributions of
the dots over the different panels clearly indicate the importance of
kinematics for the distributions of $f1$ and $f2$.  In particular,
nearly all of the points with $f1>0.3$ (representing $p/k >
5000-10^4\,{\rm cm}^{-3}\,$K) arise from gas outside the allowed
velocity bin.  The points in panel (d) representing $p/k\sim
10^5-10^6\,{\rm cm}^{-3}\,$K (with $f2>0.3$) all come from an isolated,
negative velocity component in front of $\lambda$~Cep and probably
represent gas compressed by the stellar wind from the star (and possibly
others in its association).

It is important to note that most of the C~I measurements do not fall on
the tracks, but rather just above them.  Adopting the ``center of mass''
interpretation discussed earlier, we explain this outcome in terms of
the low pressure gas (in the range $10^3 < p/k < 10^4\,{\rm cm}^{-3}$K)
being accompanied nearly always by a small amount of much higher
pressure material at $p/k > 10^5\,{\rm cm}^{-3}\,$K.  This seems to
happen for all three categories of velocity.  In the absence of this
small high-pressure contribution, we obtained the fractions at different
pressures given in Table~1.  The median thermal pressure for the entire
sample was $p/k=2240\,{\rm cm}^{-3}\,$K for $T\sim 40\,$K.

\begin{center}
\begin{tabular}{l c c c c}
\multicolumn{5}{l}{{\bf Table~1.} Distribution of Pressures } \\
\hline
\multicolumn{1}{c}{$p/k {\rm (cm}^{-3}$K)}&\multicolumn{1}{c}{$<10^3$}
&\multicolumn{1}{c}{$10^{3.0}-10^{3.5}$}
&\multicolumn{1}{c}{$10^{3.5}-10^{4.0}$}
&\multicolumn{1}{c}{$>10^{4.0}$}\\
\hline
C~I fraction & 0.028 & 0.650 & 0.285 & 0.037\\
\hline
\end{tabular}
\end{center}

\section{Possible Origins for the Very High Pressure Gas}

There is considerable evidence that supports the existence of gas
concentrations that have small angular scales (see the article
by Lauroesch, this volume).  Heiles [3] argues
that these structures are cold and probably represent sheets of gas seen
edge-on.  However, the position of the $T=20\,K$ track shown in
Fig.~\ref{f1f2_diag}$(a)$ indicates that low-temperature gas cannot be
responsible for raising the points above the tracks.  Instead, the
high-pressure material has $T\geq 40\,$K.  One possible origin is the
existence of thin, compressed zones at the outer edges of clouds that
are moving through an intercloud medium.  These zones might be
pressurized by the Bernoulli effect and heated by the dissipation of
waves arising from the Kelvin-Helmholtz instabilities, and they might
explain the production of CH$^+$ in the interstellar medium [4].

Over recent years, there has been substantial activity in modeling
interstellar turbulence and its influences on the physical state of the
medium [5].  In a medium dominated by
supersonic turbulence, the small amounts of high-pressure C~I could
arise from short-lived sheets of gas compressed by the inertial forces
of colliding gas flows.  We argue that the time scales for this
compression must be much shorter than the time the gas could reach
thermal equilibrium by various radiative processes, otherwise it would
cool too much.  Put differently, the barytropic index $\gamma_{\rm eff}$
must be intermediate between 0.72 for thermal equilibrium [6]
and $c_p/c_v=5/3$ for adiabatic compression.  This could
happen if the compressed zones had characteristic thicknesses $<
0.1\,$pc.  The compression time scales are about 4 orders of magnitude
shorter than the time needed to form appreciable amounts of H$_2$, thereby
alleviating objections stated by Heiles [3] on the overproduction of molecules.

\acknowledgements{This research was supported by NASA Grant NAS5-30110
to Princeton University and is based on observations with the NASA/ESA
Hubble Space Telescope obtained at the Space Telescope Science
Institute, which is operated by the Association of Universities for
Research in Astronomy, Incorporated, under NASA contract NAS5-26555.  A
detailed presentation of this research appears in [2].}

\vfill
\end{document}